\documentclass[useAMS, usenatbib]{mn2e} \usepackage{rotating,
graphicx, moreverb, lscape, epsfig} \def\simless{\mathbin{\lower
3pt\hbox {$\rlap{\raise 4pt\hbox{$\char'074$}}\mathchar"7218$}}}
\author[G. Raymond et al.]{G. Raymond$^1$\thanks{E-mail:
gwenifer.raymond@astro.cf.ac.uk}, S. A. Eales$^1$, S. Dye$^1$,
R. Carlberg$^2$ and M. Sullivan$^3$ \\ \\ $^1$Cardiff University,
School of Physics \& Astronomy, Queens Buildings, The Parade, Cardiff,
CF24 3AA, U.K. \\ $^2$Department of Astronomy and Astrophysics,
University of Toronto, Toronto, ON M5S 3H4, Canada\\ $^3$Department of
Physics (Astrophysics), University of Oxford, Denys Wilkinson
Building, Keble Road, Oxford, OX1 3RH, U.K.}  \title[Is there a
redshift cutoff for submillimetre galaxies?]{Is there a redshift
cutoff for submillimetre galaxies?}

\begin{document}
\date{27-07-09}

\pagerange{\pageref{firstpage}--\pageref{lastpage}} \pubyear{Submitted}

\maketitle

\label{firstpage}

\begin{abstract}
We present new optical and infrared photometry for a statistically
complete sample of seven 1.1 mm selected sources with accurate
coordinates.  We determine photometric redshifts for four of the seven
sources of 4.64, 4.54, 1.49 and 0.18.  Of the other three sources two
are undetected at optical wavelengths down to the limits of very deep
Subaru and Canada-France-Hawaii Telescope images ($\sim$27 mag AB, i
band) and the remaining source is obscured by a bright nearby galaxy.
The sources with the highest redshifts are at higher redshifts than
all but one of the $\sim$200 sources taken from the largest recent 850
$\mu$m surveys, which may indicate that 1.1 mm surveys are more
efficient at finding sources at very high redshifts than 850 $\mu$m
surveys.

We investigate the evolution of the number density with redshift of
our sample using a banded $V_{e}/V_{a}$ analysis and find no evidence
for a redshift cutoff, although the number of sources is very small.
We also perform the same analysis on a statistically complete sample
of 38 galaxies selected at 850$\mu$m from the GOODS-N field and find
evidence for a drop-off in the number density beyond $z\sim1$ and 2,
confirming the earlier conclusion of Wall, Pope \& Scott (2008).  We
also find evidence for the existence of two differently evolving
sub-populations separated in luminosity, with the drop-off in density
for the low-luminosity sources occurring at a lower redshift.
\end{abstract}


\section{Introduction}
Submillimetre (submm) galaxies (SMGs), first detected at 850 $\mu$m
with the Submillimetre Common User Bolometer Array (SCUBA)
\citep{holland:1999}, are a significant population of high redshift
star forming galaxies \citep[e.g.,][]{hughes:1998,blain:2002}.  They
are believed to be dust enshrouded galaxies undergoing prodigious
levels of star formation \citep[e.g.,][]{hughes:1998,eales:1999} in
which the optical/UV radiation emitted by the stars is absorbed by the
dust and then re-emitted in the submm.  Star formation rates in excess
of 1000 M$_{\odot}$yr$^{-1}$ have been inferred \citep{scott:2002},
much higher than locally.  The galaxies in these samples have been
found to account for up to one tenth of the total far-infrared/submm
energetic background \citep[e.g.,][]{dye:2007} and many authors have
argued that these galaxies are the progenitors for the elliptical
galaxies we see in the local Universe
\citep{eales:1999,scott:2002,dunne:2003}.  Thus understanding the
nature of these sources is of great importance for the understanding
of galaxy formation and evolution as a whole.

Observations of SMGs at $\sim$1 mm benefit from a negative
K-correction out to high redshifts due the shape of their spectral
energy distribution (SED).  As the redshift of an SMG increases, the
peak of its rest-frame SED moves toward the observed waveband,
offsetting the dimming caused by the increasing luminosity distance.
This fact accounts for the surprising ability of SCUBA to find large
numbers of high-redshift galaxies.

The large amount of dust responsible for the strong submm emission
gives rise to high levels of attenuation in the optical.  This in
conjunction with the poor angular resolution of single dish submm
facilities makes the cross identification of SMGs at different
wavelengths difficult.  Moreover, even when an optical counterpart can
be identified, the high levels of dust attenuation makes the
determination of a spectroscopic redshift difficult.  As such we are
currently unable to determine spectroscopic redshifts for the majority
of SMGs.  The strong correlation between dust emission and radio
emission which appears to hold true in both the low-redshift and
high-redshift universe \citep{vlahakis:2007} has been useful for both
identifying the counterparts and estimating redshifts.  Due to the low
surface density of radio sources on the sky, the probability of the
radio counterpart being coincidental with the submm source by chance
is small.  Due to the high positional accuracy of radio observations,
it is then possible to identify the optical counterpart and retrieve a
spectroscopic redshift.  It is also possible to estimate the redshift
using the ratio of radio to submm flux \citep[e.g.,][]{hughes:1998,
carilli:1999, carilli:2000, smail:2000}.

\citet{chapman:2005}, using the Low Resolution Imaging Spectrograph
(LRIS) \citep{oke:1995} on the Keck I telescope, managed to obtain
spectroscopic redshifts for a total of 73 radio-identified SMGs with a
median 850 $\mu$m flux of 5.3 mJy.  The galaxies in this sample were
found to lie at a median redshift of $z=2.2$ out to a maximum value of
$z_{max}=3.6$.  However, the K-correction which allows us to detect
high-redshift SMGs does not similarly benefit their radio fluxes and
so radio identified SMGs are subjected to a radio selection effect
which limits redshifts to $z\simless3$.

\citet{pope:2006}, produced the first complete (i.e. not requiring
radio IDs) sample of 850 $\mu$m selected SMGs that has close to 100\%
redshifts.  The sample consists of 35 galaxies, 21 with secure optical
counterparts and 12 with tentative optical counterparts, and its
completeness means that unlike previous surveys it is not biased
towards low-z sources.  The median redshift determined for this sample
is $z\sim 2.2$.  Using this sample, \citet{wall:2008} examined the
epoch dependency of the number density of SMGs.  They found an
apparent redshift cutoff at $z>3$ with further evidence for two
separately evolving populations, divided by luminosity.  However this
result was based on calculations using a single model galaxy SED.
Since the predicted relationship between submm flux-density and
redshift depends strongly on the assumed SED, one of the aims of this
paper is to re-examine their conclusion using a range of empirical
SEDs.

There have been a number of explanations for the lack of high-redshift
SMGs.  Given that dust is thought to form in the atmospheres of highly
evolved stars, it is possible that at high redshifts simply not enough
time has passed for dust to form \citep{morgan:2003}.  Observations of
high-redshift quasars have however detected high levels of dust
\citep[e.g.,][]{priddey:2001}, suggesting that this is not the
explanation.  Another possible explanation is that there are fewer
large star-forming galaxies at high redshifts.

\citet{eales:2003} presented evidence that SMGs typically have low
values for the ratio of the 850 $\mu$m to 1200 $\mu$m fluxes compared
to that expected from a low redshift galaxy.  One possible explanation
is that these sources are at very high redshifts.  If this is true,
then observations at 1.1 mm would be better at detecting SMGs at the
highest redshifts than observations at 850 $\mu$m.  A new complete
sample of 1.1 mm selected SMGs located in the COSMOS field
\citep{scoville:2007} has been compiled by \citet{younger:2007}.  The
sources were selected initially at 1.1 mm with the AzTEC camera
\citep{scott:2008, wilson:2008} on the JCMT.  The resultant catalogue
consists of 44 sources with S/N$>3.5\sigma$, 10 of which are robust
with S/N$>5\sigma$.  Follow up observations by \citet{younger:2007}
were then made with the Submillimetre Array (SMA) at 890 $\mu$m for
the 7 highest significance AzTEC sources, allowing their positions to
be determined with an accuracy of $\sim 0.2$''.  The COSMOS field
offers a wealth of data over a great number of wavebands including the
optical and infrared.  Thus the high positional accuracy allows for
the identification of optical counterparts and hence the determination
of photometric redshifts.  Of the seven AzTEC sources imaged with the
SMA six have IRAC counterparts, and one source is obscured by a nearby
bright galaxy.  Using deep Hubble Space Telescope (HST) imaging
acquired with the Advanced Camera for Surveys (ACS),
\citet{koekemoer:2007} found optical counterpart candidates for only
three of these sources.

The main aim of this paper is to carry out a deeper search for the
optical counterparts for the AzTEC sources.  We give photometry from
deep Subaru and Canada-France-Hawaii-Telescope (CFHT) imaging and find
one new possible optical ID.  We estimate photometric redshifts for
the AzTEC sources using the {\scshape HyperZ} photometric redshift
package \citep{bolzonella:2000}.  Throughout this work we employ a
concordance cosmological model with $\Omega_{total}=1$,
$\Omega_{m}=0.3$, $\Omega_{\Lambda}=0.7$ and $H_{0} = 75$
kms$^{-1}$Mpc$^{-1}$.  All magnitudes quoted are AB.

\section{NEW IMAGES AND PHOTOMETRY FOR THE AzTEC SAMPLE}

We searched for optical counterparts and measured new photometry using
deep Subaru\footnote{An additional uncertainty of 0.3 mags in the
Subaru B$_{j}$ band magnitudes is taken into account in this
photometry due to the possibility of a red leak or a shift in the blue
cutoff of this filter.}, CFHT and IRAC images of the AzTEC sources.
The IRAC and Subaru images are the publicly available COSMOS images
taken by the COSMOS team \citep{scoville:2007}.  The CFHT images are
taken from the CFHT Deep Legacy Survey.  The images we used were taken
using the CFHT $g_{\mathrm{M}}$, $r_{\mathrm{M}}$, $i_{\mathrm{M}}$,
$z_{\mathrm{M}}$, Subaru B$_{j}$, V$_{j}$, r+, i+, z+ and IRAC channel
1 and 2 filters to average 3$\sigma$ depths of approximately 28.4,
27.9, 27.6, 26.5, 29.0, 28.2, 28.3, 27.7, 26.4, 24.1 and 23.6 mags
respectively.

We searched the i-band images (figure~\ref{fig:images}) at the SMA
coordinates.  We find bright i-band counterparts for AzTEC1, 3 and 7,
all of which were previously known.  We also find a faint i-band
counterpart for AzTEC5 at the SMA coordinates.  We find no objects
directly at the SMA coordinates for AzTEC2, but there is a bright
object offset from this position by 3'', meaning that the magnitude
limits of this SMG are not useful.

We find no optical counterparts directly at the SMA coordinates for
AzTEC4 and 6 in the Subaru and CFHT imaging.  For the latter source,
however, there is a bright i-band counterpart offset from the SMA
position by $\sim$0.6'' ($\sim$3$\sigma$) which could be AzTEC6's
counterpart or the true counterpart may be too faint to see.  There is
also a faint i-band source, offset from AzTEC4's SMA position by
$\sim$0.8'' ($\sim$4$\sigma$), in the Subaru imaging.  For these two
sources we added the i band CFHT and Subaru images, inversely
weighting the images by the square of the noise, in order to try and
detect any very faint possible counterparts.  The stacked i-band
images for AzTEC4 and 6 are shown in figure~\ref{fig:images}.  We
still do not find counterparts at the SMA positions for AzTEC4 and 6
and given the good coincidence between the SMA and optical positions
for the other AzTEC sources we tentatively conclude that the true
counterparts have not yet been detected.

The typical full-width half-maximum (FWHM) of the optical and IRAC
channel 1 and 2 point spread functions (PSFs) are $\sim$0.8'', 1.66''
and 1.72'' respectively.  Magnitudes were determined manually by
placing apertures onto the images, ensuring that the aperture was
large enough to contain as much of the emission from the galaxy as
possible without including any emission from neighboring objects.
Thus the sizes of the optical apertures vary from source to source,
although are constant for a given source.  We used larger apertures
for the IRAC sources due to the images having a larger PSF, but use
the procedure outlined below to correct for this.

Due to the difference in the PSF between the optical and IRAC images
as well as the difference in the aperture sizes, a small correction
needed to be applied to the IRAC magnitudes before they could be used
in conjunction with optical magnitudes to determine a photometric
redshift.  We corrected IRAC magnitudes by firstly fitting a 2D
Gaussian to the IRAC source.  We then scaled it to have the FWHM it
would have had if observed with CFHT/Subaru.  The flux was then
computed using the scaled Gaussian and new aperture size.  All the
corrections applied in this work increase the IRAC magnitudes, and the
more extended the source the greater the correction.  Corrections
range from 0.01 to 0.46 magnitudes.

The new photometry is summarized in table~\ref{table:magnitudes}, and
the details of the individual objects are discussed below.

\begin{figure*}
\epsfxsize=170mm
{\hfill
\epsfbox{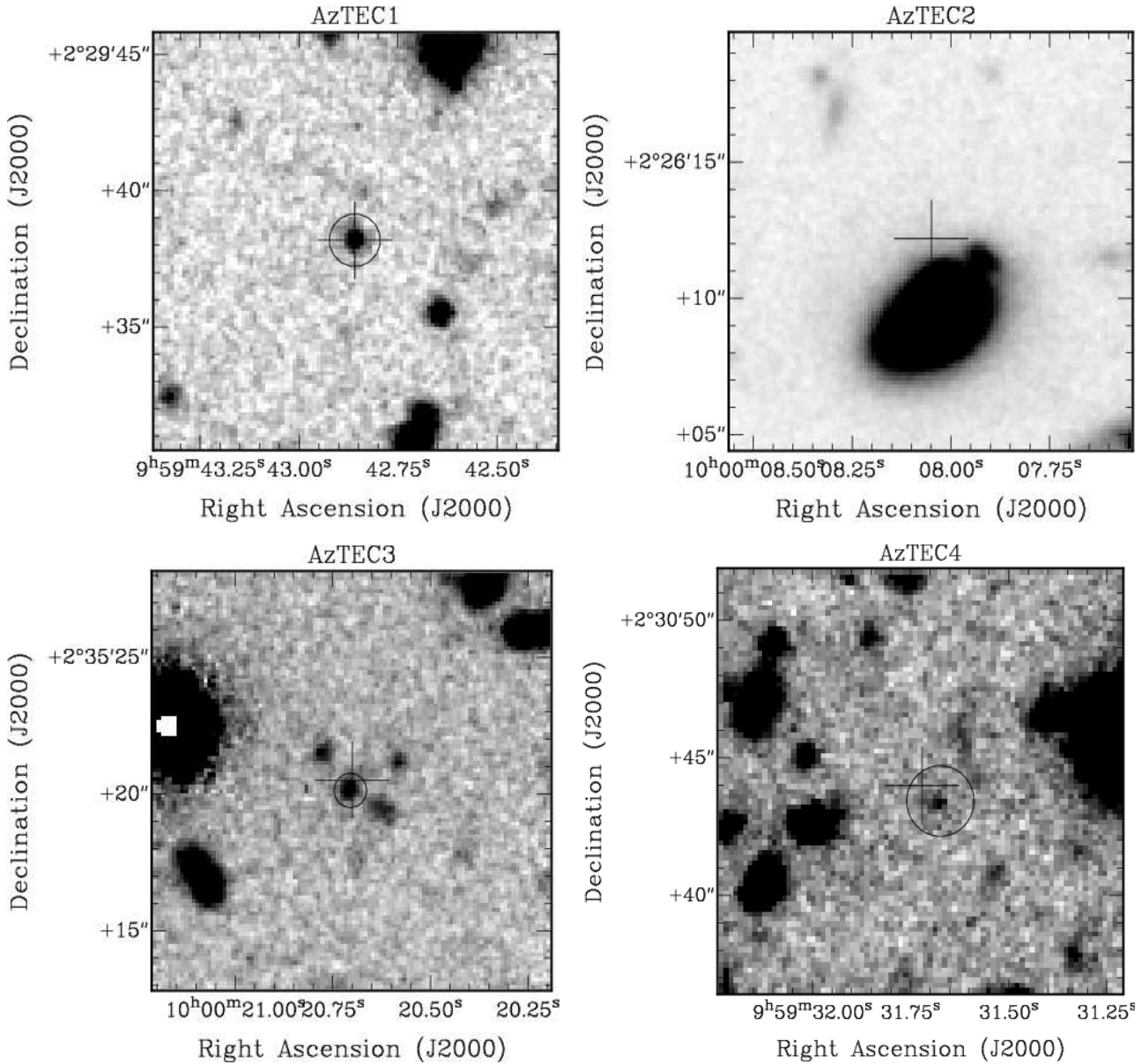}
\hfill}
  \caption{Subaru i+ band cutouts for AzTEC1 to 7, with the exception
  of AzTEC4 and 6, which are the combined CFHT and Subaru i-band
  images.  Each image has a field of view of 15.3''x15.3'' and a scale
  of 0.15''/pixel.  The SMA coordinates of each source are highlighted
  by cross-hairs and the optical counterpart (including the objects
  offset from AzTEC4' and 6s SMA coordinates by $\sim$0.8'' and
  $\sim$0.6'' respectively, see text) is circled with an aperture the
  size of which was used in its photometry.  SMA coordinates are
  accurate to $\sim$0.2''.}
  \label{fig:images}
\end{figure*}

\begin{figure*}
\epsfxsize=170mm
{\hfill
\epsfbox{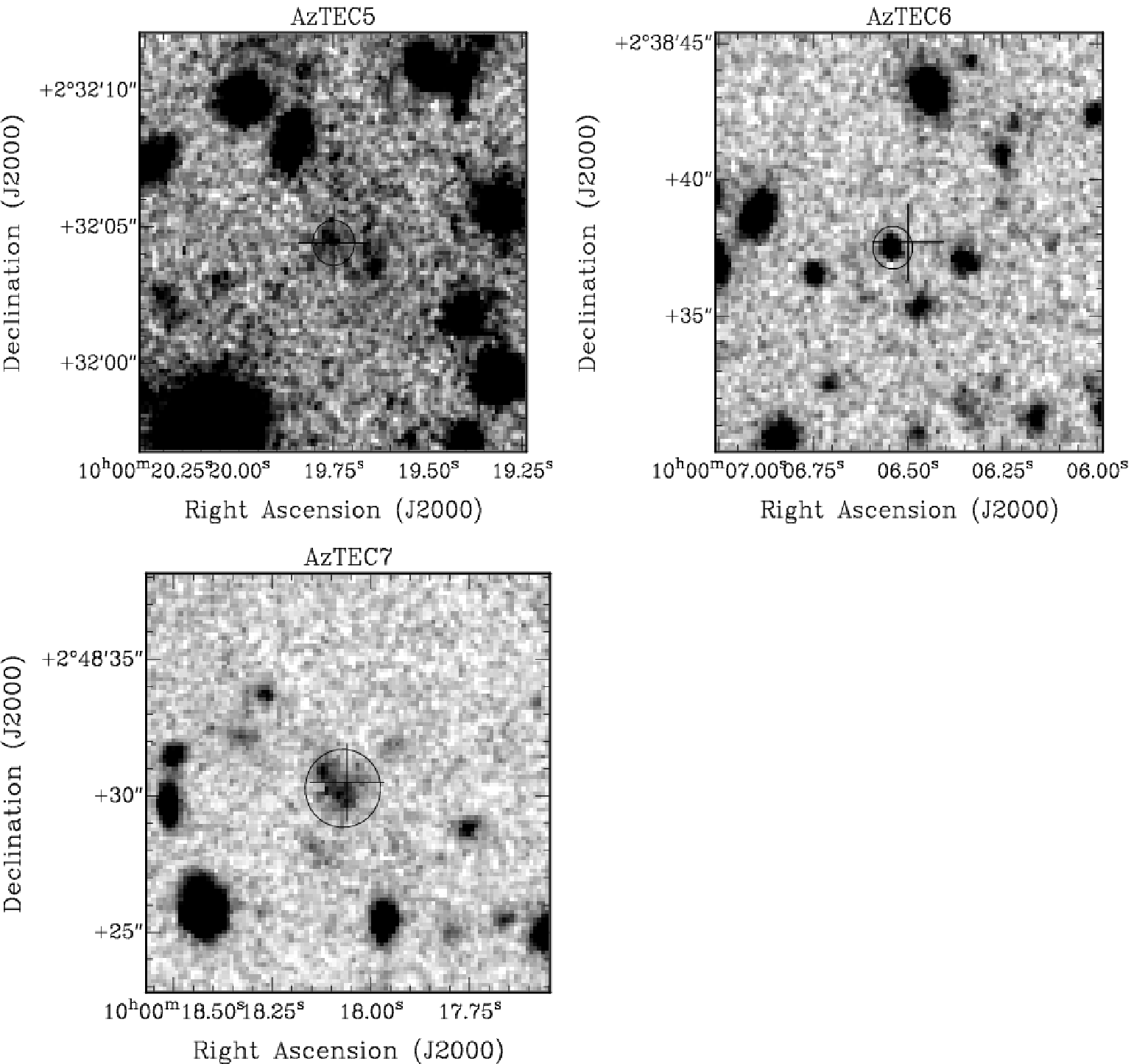}
\hfill}
  \contcaption{}
 \label{fig:images}
\end{figure*}

\subsection{Notes on Individual Objects}

\noindent\bf{AzTEC1}\rm - \it{J095942.86+022938.2}\rm - AzTEC1 is the
brightest submm source in the sample with fluxes of $F_{890\mu m} =
15.6\pm1.1$ mJy and $F_{1.1mm} = 10.7\pm1.3$ mJy.  There is a bright
i-band object located directly at the SMA position.  Optical fluxes
are measured using an aperture 1.94'' in diameter and the source is
detected in the Subaru i+ band at 25.11$\pm$0.03 mag which is in
agreement with the HST i-band magnitude given in \citet{younger:2007}.
There is some disagreement within the same wavebands between the CFHT
and Subaru and photometry (table~\ref{table:magnitudes}), but the
discrepancy is small enough that the photometric redshift is not
significantly affected.  IRAC magnitudes were measured using an
aperture of diameter 4.45''.  Only a small correction was applied to
the IRAC magnitudes: +0.01 mag in both IRAC channel 1 and 2.
\newline
\newline
\noindent\bf{AzTEC2}\rm - \it{J100008.05+022612.2}\rm - AzTEC2 is
detected in the submm with fluxes $F_{890\mu m} = (12.4\pm1.0)$ mJy
and $F_{1.1mm} = (9.0\pm1.3)$ mJy.  No objects are found directly at
the SMA coordinates, but there is a bright object offset from SMA
position by 3''.  Thus the limit on the magnitude of the optical
counterpart is not very useful.
\newline
\newline
\noindent\bf{AzTEC3}\rm - \it{J100020.70+023520.5}\rm - AzTEC3 is
detected in the submm with fluxes $F_{890\mu m} = 8.7\pm1.5$ mJy and
$F_{1.1mm} = 7.6\pm1.2$ mJy.  There is a bright i-band object located
at the SMA coordinates as well as three companion objects offset by
between 1'' and 2''.  Since SMGs often seem to consist of multiple
components \citep{ivison:1998} it is possible that these companion
objects are also part of AzTEC3.  However, since six of the seven
AzTEC sources are detected in the IRAC bands, it seems likely that if
the companion objects are part of the same galaxy then they should
also be contributing to the IRAC emission.  We attempted to determine
whether this is the case by convolving the Subaru image with the IRAC
beam and comparing the FWHM of the IRAC source with that of the
convolved Subaru image.  We find that the FWHM of the convolved image
is $\sim$4.4''.  The FWHM of the IRAC 3.6 $\mu$m image is
$\sim$2.87'', suggesting that the 3.6 $\mu$m emission is associated
only with the central object.  Optical fluxes were measured using a
aperture of diameter 1.26'' and the source is detected in the Subaru
i+ band at 26.18$\pm$0.08 mag which is in agreement with the HST
i-band magnitude.  The CFHT and Subaru magnitudes within the same
bands are consistent with each other.  IRAC magnitudes were measured
using an aperture of diameter 4.80''.  A correction of +0.2 mag was
applied to the IRAC magnitudes in channels 1 and 2.
\newline
\newline
\noindent\bf{AzTEC4}\rm - \it{J095931.72+023044.0}\rm - AzTEC4 is
detected in the submm with fluxes $F_{890\mu m} = 14.4\pm1.9$ mJy and
$F_{1.1mm} = 6.8\pm1.3$ mJy.  We find a tentative i-band counterpart,
offset from the SMA position by 0.8'' ($\sim$3$\sigma$), in the Subaru
image with a magnitude of 27.43$\pm$0.13.  In the combined image (see
above), the counterpart can be seen more clearly and has a magnitude
of 26.99$\pm$0.18 in a 2.57'' diameter aperture.  However we find it
is too faint to detect in our other Subaru and CFHT bands.  IRAC
magnitudes were measured using an aperture of diameter 4.80''.
Corrections of +0.11 and +0.04 mags were applied to the IRAC
magnitudes in channels 1 and 2 respectively.  Because of the good
agreement between the SMA the optical positions for the other AzTEC
sources we tentatively conclude that this is not the true counterpart.
\newline
\newline
\noindent\bf{AzTEC5}\rm - \it{J100019.75+023204.4}\rm - AzTEC5 is
detected in the submm with fluxes $F_{890\mu m} = 9.3\pm1.3$ mJy and
$F_{1.1mm} = 7.6\pm1.3$ mJy.  \citet{younger:2007} found no optical
counterpart in ACS imaging, but we find a faint Subaru source at the
SMA coordinates with a Subaru i+ band magnitude of 26.74$\pm$0.13,
measured in an aperture of diameter 1.68''.  The CFHT and Subaru
magnitudes within the same bands are consistent with each other.  IRAC
magnitudes were measured using a aperture of diameter 4.80''.
Corrections of +0.46 and +0.14 mag were applied to the IRAC magnitudes
in channels 1 and 2 respectively.
\newline.
\newline
\noindent\bf{AzTEC6}\rm - \it{J100006.50+023837.7}\rm - AzTEC6 is
detected in the submm with fluxes $F_{890\mu m} = 8.6\pm1.3$ mJy and
$F_{1.1mm} = 7.9\pm1.2$ mJy.  \citet{younger:2007} find no optical
counterpart in ACS imaging.  In CFHT and Subaru imaging we find no
source directly at the SMA coordinates, but we do find a source offset
from the SMA position by $\sim$0.6'' ($\sim$3$\sigma$).  This could
therefore be the optical counterpart, or the true counterpart may be
too faint to detect.  The source offset from the SMA position has a
Subaru i+ magnitude of 25.38$\pm$0.04 magnitudes, measured in an
aperture of diameter 1.59''.  The CFHT and Subaru magnitudes within
the same bands are consistent with each other.  IRAC magnitudes were
measured using an aperture of diameter 5.88''.  A correction of +0.13
mag is applied to the IRAC magnitudes in channels 1 and 2.  Because of
the good agreement between the SMA the optical positions for the other
AzTEC sources we tentatively conclude that this is not the true
counterpart, although we do estimate a photometric redshift for it.
\newline
\newline
\noindent\bf{AzTEC7}\rm - \it{J100018.06+024830.5}\rm - AzTEC7 is
detected in the submm with fluxes $F_{890\mu m} = 12.0\pm1.5$ mJy and
$F_{1.1mm} = 8.3\pm1.4$ mJy.  We find an optical counterpart with a
disturbed morphology at the SMA coordinates which could be a system of
merging galaxies.  Optical fluxes were measured by placing an aperture
of diameter 2.87'' over the whole of the system.  The source is
detected in the Subaru i+ band at 24.20$\pm$0.04 mag.  IRAC magnitudes
were measured using an aperture of diameter 6.12''.  Corrections of
+0.08 and +0.02 mag were applied to the IRAC magnitudes in channels 1
and 2 respectively.
\newline.
\newline

\begin{table*}
     \centering
      \begin{tabular}{cccccccc}
      \hline
                   &        AzTEC1       &        AzTEC2       &       AzTEC3        &       AzTEC4        &      AzTEC5         &       AzTEC6           &      AzTEC 7   \\
      \hline
            RA     &     09:59:42.86     &     10:00:08.05     &     10:00:20.70     &     09:59:31.72     &    10:00:19.75      &     10:00:06.50        &    10:00:18.06 \\
            Dec    &     +02:29:38.2     &     +02:26:12.2     &     +02:35:20.5     &     +02:30:44.0     &    +02:32:04.4      &     +02:38:37.7        &    +02:48:30.5 \\
  Optical Ap. Size &        1.94''       &         ...         &        1.26''       &       2.57''        &       1.68''        &     ...(1.59'')        &       2.87''    \\
      $m_{B}$      &       $>$28.88      &         ...         &       $>$29.14      &      $>$28.67       &   28.80$\pm$0.47    &$>$28.98(25.80$\pm$0.30)&   25.67$\pm$0.31\\
      $m_{V}$      &    27.13$\pm$0.17   &         ...         &    28.77$\pm$0.75   &      $>$28.21       &   28.76$\pm$0.62    &$>$28.32(25.67$\pm$0.04)&   25.10$\pm$0.06\\
      $m_{r+}$     &    26.21$\pm$0.06   &         ...         &    27.39$\pm$0.22   &      $>$28.02       &   27.07$\pm$0.13    &$>$28.47(25.77$\pm$0.03)&   24.97$\pm$0.05\\
      $m_{i+}$     &    25.11$\pm$0.03   &         ...         &    26.18$\pm$0.08   &  (27.43$\pm$0.13)   &   26.74$\pm$0.13    &$>$27.97(25.38$\pm$0.04)&   24.20$\pm$0.04\\
      $m_{z+}$     &    25.02$\pm$0.02   &         ...         &    25.58$\pm$0.15   &      $>$26.50       &   26.07$\pm$0.22    &$>$26.79(24.80$\pm$0.06)&   23.65$\pm$0.07\\
  $g_{\mathrm{M}}$ &       $>$28.12      &         ...         &       $>$28.71      &      $>$27.93       &      $>$28.41       &$>$28.42(26.16$\pm$0.05)&         N/A     \\
  $r_{\mathrm{M}}$ &    26.54$\pm$0.15   &         ...         &    27.13$\pm$0.24   &      $>$27.46       &   27.15$\pm$0.20    &$>$27.96(25.61$\pm$0.05)&         N/A     \\
  $i_{\mathrm{M}}$ &    25.25$\pm$0.05   &         ...         &    26.30$\pm$0.12   &      $>$26.19       &   26.50$\pm$0.13    &$>$27.81(25.42$\pm$0.05)&         N/A     \\
  $z_{\mathrm{M}}$ &    25.11$\pm$0.13   &         ...         &    25.69$\pm$0.22   &      $>$26.27       &   26.46$\pm$0.38    &$>$26.06(24.87$\pm$0.08)&         N/A     \\
   IRAC Ap. Size   &         4.45''      &         ...         &         4.80''      &       4.80''        &        4.80''       &        5.88''          &        6.12''   \\
   $m_{3.6\mu m}$  &    23.40$\pm$0.07   &         ...         &    23.72$\pm$0.11   &    22.11$\pm$0.04   &   23.24$\pm$0.08    &    24.13$\pm$0.25      &   20.63$\pm$0.01\\
   $m_{4.5\mu m}$  &    23.08$\pm$0.08   &         ...         &    22.98$\pm$0.12   &    22.15$\pm$0.04   &   22.31$\pm$0.06    &    23.50$\pm$0.27      &   20.15$\pm$0.02\\
     \end{tabular}
    \caption{Photometry for the AzTEC sources, given in AB magnitudes.
First two rows give the SMA co-ordinates.  Aperture sizes are the
diameters used for measuring optical and IRAC magnitudes.  The IRAC
magnitudes are corrected to take into account the difference in the
seeing and aperture sizes for the IRAC and optical imaging (see text).
No optical counterparts were found for AzTEC2.  The only nearby
optical counterparts for AzTEC4 and 6 are offset from their SMA
positions by $\sim$4$\sigma$ and $\sim$3$\sigma$ respectively.  We
give the photometry for these objects in parentheses.}
    \label{table:magnitudes}
\end{table*}

\section{ESTIMATED REDSHIFTS FOR THE AzTEC SAMPLE}
Photometric redshifts were determined by applying the photometric
redshift package, {\scshape HyperZ} \citep{bolzonella:2000}, to our 11
band photometry (Subaru:B, V, r+, i+, z+; CFHT: $g_{\mathrm{M}}$,
$r_{\mathrm{M}}$, $i_{\mathrm{M}}$, $z_{\mathrm{M}}$; IRAC: 3.6
$\mu$m, 4.5 $\mu$m).  The spectra used for fitting in this work are
taken from the set compiled by \citet{dye:2008}, which is optimized
for the determination of photometric redshifts when including filters
in the near/mid-infrared.  \citet{dye:2008} compared the photometric
redshifts determined using these spectral templates with those
determined using synthetic spectra constructed from the best-fit star
formation history for their sample of 60 SCUBA sources.  Since these
methods are completely independent and the redshifts found using both
sets of templates were found to be in good agreement, we assume that
our template set is adequate.

We varied the redshift in the range $z = 0$ to 10.  We employed the
reddening regime of \citet{calzetti:2000}, with $A_{V}$ allowed to
vary in the range $A_{V}= 0$ to 5 in steps of 0.2.  We used a minimum
photometric error of 0.05 magnitudes for each band.  For wavebands in
which we have no detection we took the flux of the source to be zero
with a $1\sigma$ error equal to the sensitivity of the detector in
that waveband.  The photometric redshifts obtained are listed in
table~\ref{table:redshift}.

\begin{table*}
  \centering
  \begin{tabular}{cccc}
   \hline ID & z & $\chi^{2}_{min} $ & Notes \\
   \hline  AzTEC1 & $4.64\pm0.06$ & 1.537 & ...\\
           AzTEC2 & ... & ... &             No optical counterpart.\\
	   AzTEC3 & $4.54\pm0.10$ & 2.196 & There is a secondary chi-squared minimum at the lower \\ 
                      &&&                   redshift of $z\sim0.4$ with a chi-squared fit value of $\sim$3.5.\\
	   AzTEC4 & ... & ... &             Nearest counterpart only detected in one optical band. \\
	   AzTEC5 & $1.49\pm0.10$ & 1.488 & There is a secondary chi-squared minimum at the higher \\
                      &&&                   redshift of $z\sim4$ with a chi-squared fit value of $\sim$4. \\
	   AzTEC6 &($2.09\pm0.01$)&(6.172)& The redshift and chi-squared values are for the optical source \\
                      &&&                   offset from AzTEC6's SMA position.  The chi-squared fit to \\
                      &&&                   this source is much poorer compared to the others in the sample.\\
                      &&&                   This may further imply that the nearby optical counterpart we have \\
                      &&&                   selected is not the true counterpart to AzTEC6 and that the IRAC \\
                      &&&                   emission is unassociated with the optical emission.\\
	   AzTEC7 & $0.18\pm0.01$ & 7.021 & CFHT data not available.  There are several other possible redshifts\\ 
                      &&&                   with chi-squared fit values of $\sim10$ up to $z\sim2$.  Even the best\\
                      &&&                   chi-squared fit is still relatively poor however, which may be due\\
                      &&&                   to the unusual nature of the source.\\                                         
  \end{tabular}
  \caption{The best photometric redshift fits for the sources with
  their minimum reduced $\chi^{2}$ value, $\chi^{2}_{min}$.  Notes of
  interest on the photometric redshifts, including secondary fits, for
  each source are also given.  We do not give the best fit SED type as
  typically for each source there are several SED types which fit
  equally well.  Note that reduced chi-squared values given here are
  not those directly output by {\scshape HyperZ}, which takes the
  number of degrees of freedom as being the (number of filters$-1$).
  This is true if only the redshift is allowed to vary.  However, we
  are additionally allowing $A_{v}$, SEDs type and the normalization
  to vary.  Thus the correct number of degrees of freedom is given by
  (number of filters$-4$).}
  \label{table:redshift}
\end{table*}

\begin{figure*}
\epsfxsize=150mm
{\hfill
\epsfbox{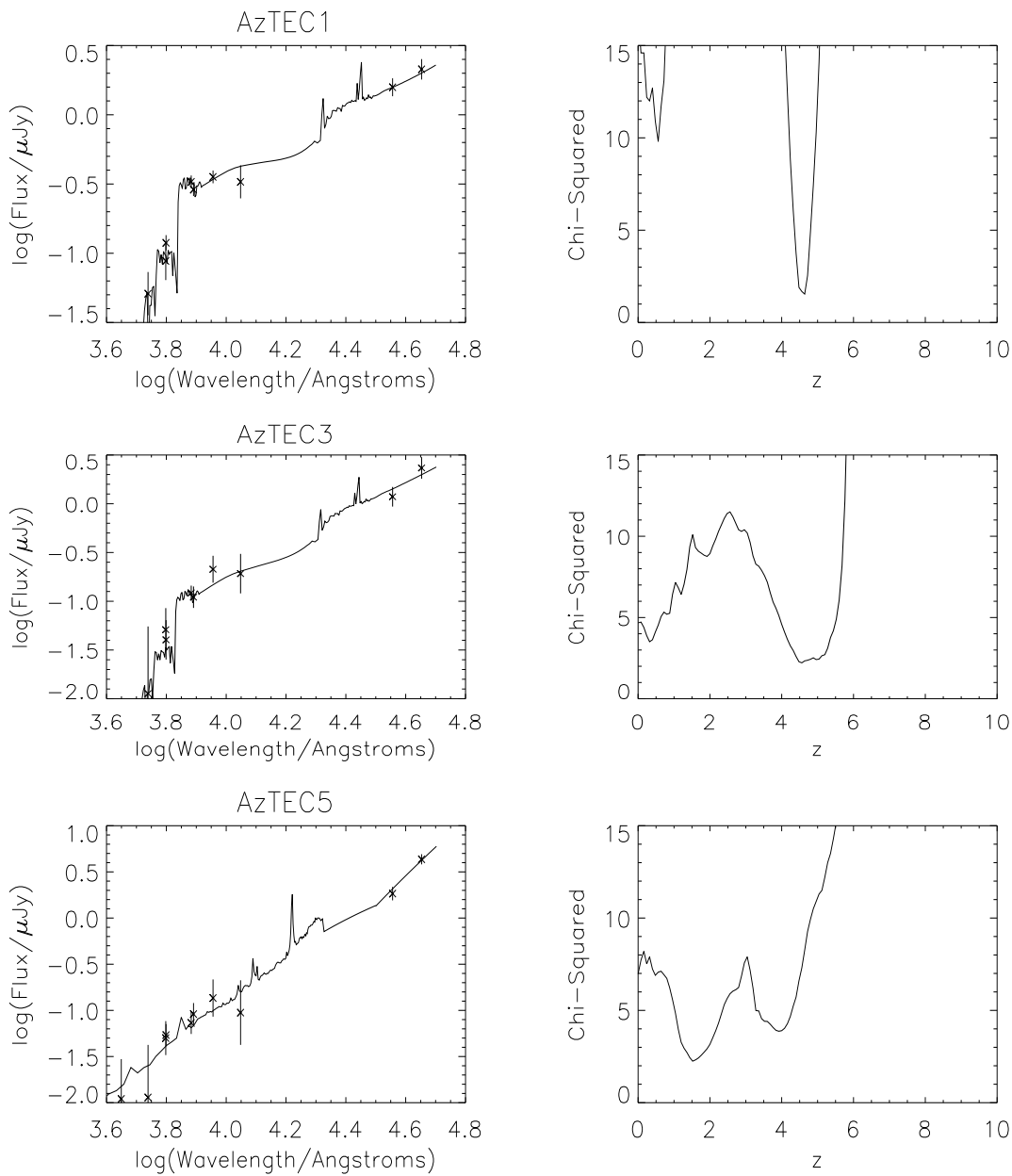}
\hfill}
  \caption{The left hand column shows the photometric data points for
  the AzTEC sources with optical counterparts.  The best spectral fits
  for the sources are overlaid.  The right hand column shows the
  marginalized reduced $\chi^{2}$ distribution as a function of
  redshift.  The AzTEC6 plots correspond to the nearby optically
  bright object.}
 \label{fig:zphot}
\end{figure*}

\begin{figure*}
\epsfxsize=150mm
{\hfill
\epsfbox{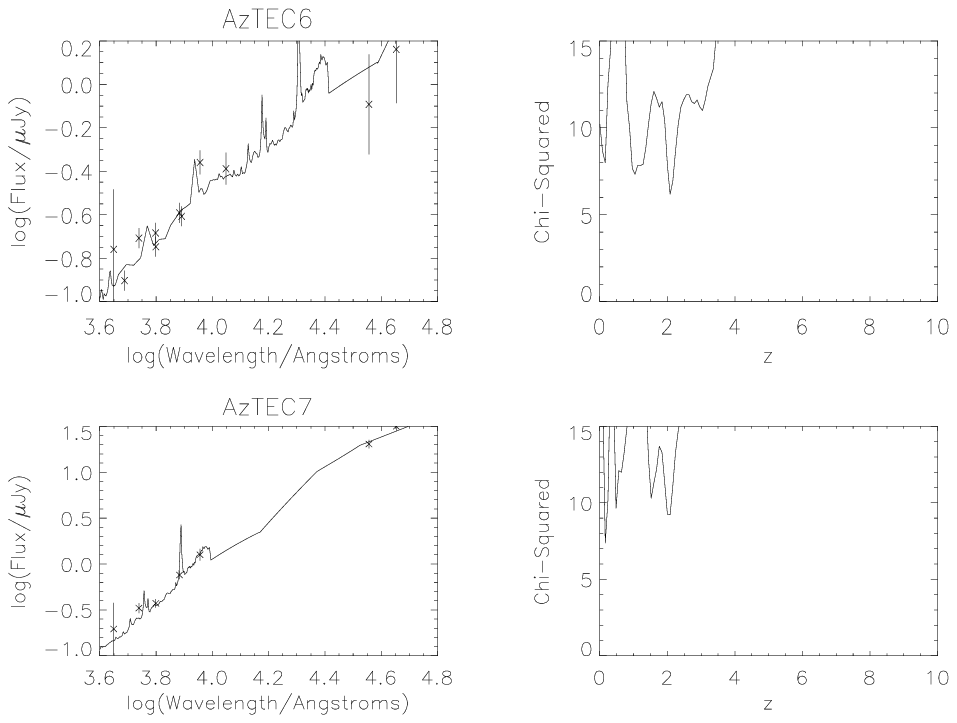}
\hfill}
  \contcaption{}
 \label{fig:zphot}
\end{figure*}

The median redshift of the sample is 2.7 which is somewhat higher than
the median redshift, 2.2, of the sample presented by
\citet{chapman:2005}.  The maximum redshift found is 4.64 and the
minimum redshift found is 0.18.  Comparing the redshift distribution
of this sample to that of the samples presented in
\citet{chapman:2005}, \citet{pope:2006}, \citet{dye:2008} and
\citet{clements:2008}, we note that only one of the sources in this
combined sample of $\sim$200 850 $\mu$m selected sources is at a
comparably high redshift as our two highest redshift sources, although
this difference is not significant when analyzed with a
Kolmogorov-Smirnov test.  However two of the other AzTEC sources are
undetected to very faint limits in the i-band, and these facts may
indicate that 1.1 mm surveys find more sources at very high redshifts
than 850 $\mu$m surveys.

\section{A BANDED $V_{e}$/$V_{a}$ ANALYSIS}

\citet{wall:2008} examined a sample of 38 SMGs in the GOODS-N field
and found evidence for a diminution in the space density of SMGs at
redshifts $z>3$.  They also found evidence for two separately evolving
sub-populations separated by luminosity.  In this paper we present the
results of our re-examination of this result using a banded
$V_{e}/V_{a}$ analysis and a range of empirical SEDs rather than the
theoretical SED used by Wall et al.

The most well known method of investigating the evolution of the space
density of galaxies with redshift is the $\langle V/V_{max}\rangle$
test \citep{schmidt:1968,rowan-robinson:1968}.  $V$ is the co-moving
volume enclosed by the galaxy (that volume which the field of view
traces out in moving from a redshift of $z=0$ out to the galaxy) and
$V_{max}$ is the volume that would be enclosed by the galaxy were it
pushed to the redshift at which its flux drops to the survey limit.
This method encounters problems when a survey encloses two galaxy
populations, one undergoing positive evolution, and the other
negative.  If we have a uniform distribution of galaxies in space,
then we expect the value of $\langle V/V_{max}\rangle$ to be
0.5$\pm(12N)^{-0.5}$, where $N$ is the number of sources in the
sample.  A value of $\langle V/V_{max}\rangle > 0.5$ then implies a
concentration of sources toward the more distant regions of their
accessible volume and a value of $\langle V/V_{max}\rangle < 0.5$
implies a deficit of sources at higher redshifts.  Therefore if we
have in our sample separate populations undergoing high levels of
positive and negative evolution, then $\langle V/V_{max}\rangle$ may
still be close to 0.5, incorrectly implying zero evolution.

This problem can be solved by implementing instead a $\langle
V_{e}/V_{a} \rangle$ test \citep{dunlop:peacock}.  This is effectively
a banded version of the $\langle V/V_{max}\rangle$ test.  $V_{e}$, the
effective volume, is the volume enclosed between a minimum redshift
$z_{low}$ and the redshift of the galaxy.  $V_{a}$, the accessible
volume, is the volume enclosed between $z_{low}$ and the redshift at
which the galaxy's flux drops below the sensitivity of the survey.  By
investigating the variation of $\langle V_{e}/V_{a}\rangle$ with
$z_{low}$ we can distinguish between a positively evolving and a
negatively evolving population.

\begin{figure}
\epsfxsize=75mm
{\hfill
\epsfbox{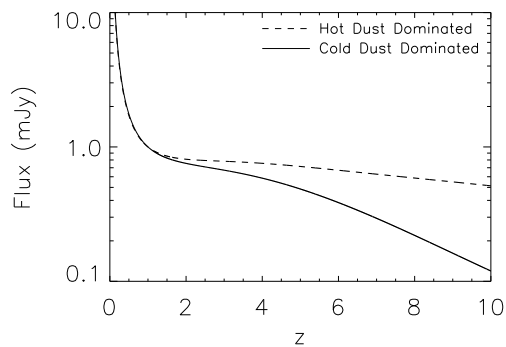}
\hfill}
  \caption{Flux versus redshift for both the cold (solid line) and hot
  (dashed line) SEDs.  Both SEDs are normalized such that they produce
  a flux of 1 mJy at a redshift of $z=1$.}
  \label{fig:f_v_z}
\end{figure}

We investigated the evolution of the space density of the sample with
redshift through the implementation of a $\langle V_{e}/V_{a} \rangle$
test.  Wall et al. based the k-correction necessary to calculate
accessible volume on a single theoretical SED, whereas real galaxies
have a range of SEDs.  To investigate this, we carried out the
$\langle V_{e}/V_{a} \rangle$ analysis using two different assumptions
about SEDs.  We used the two extreme two-component dust models of
\citet{dunne:eales}, who provided fits to the hottest and coldest
local SMGs.  The cold SED, based on NGC 958, contains dust at
temperatures of 20 and 44 K with a cold-to-hot dust mass ratio of
186:1.  The hot SED, based on IR1525+36, contains dust at temperatures
of 19 and 45 K with a cold-to-hot dust mass ratio of 15:1.
Figure~\ref{fig:f_v_z}, which shows the predicted flux versus redshift
plot for the different models, shows the effect of using different SED
templates on the flux-redshift relation.  The two SED types are
normalized such that they produce a flux of 1 mJy at a redshift of
$z=1$.

We took the limiting flux of each source in the GOODS sample to be
$3.5\sigma$ and measured $\langle V_{e}/V_{a}\rangle$ for $z_{low}=0$
to 4 in steps of 0.1.  We also separated sources into two samples of
equal size according to luminosity.  In doing this we are able to
determine whether there are differences in the evolution of the two
sub-populations.

\begin{figure*}
\epsfxsize=150mm
{\hfill
\epsfbox{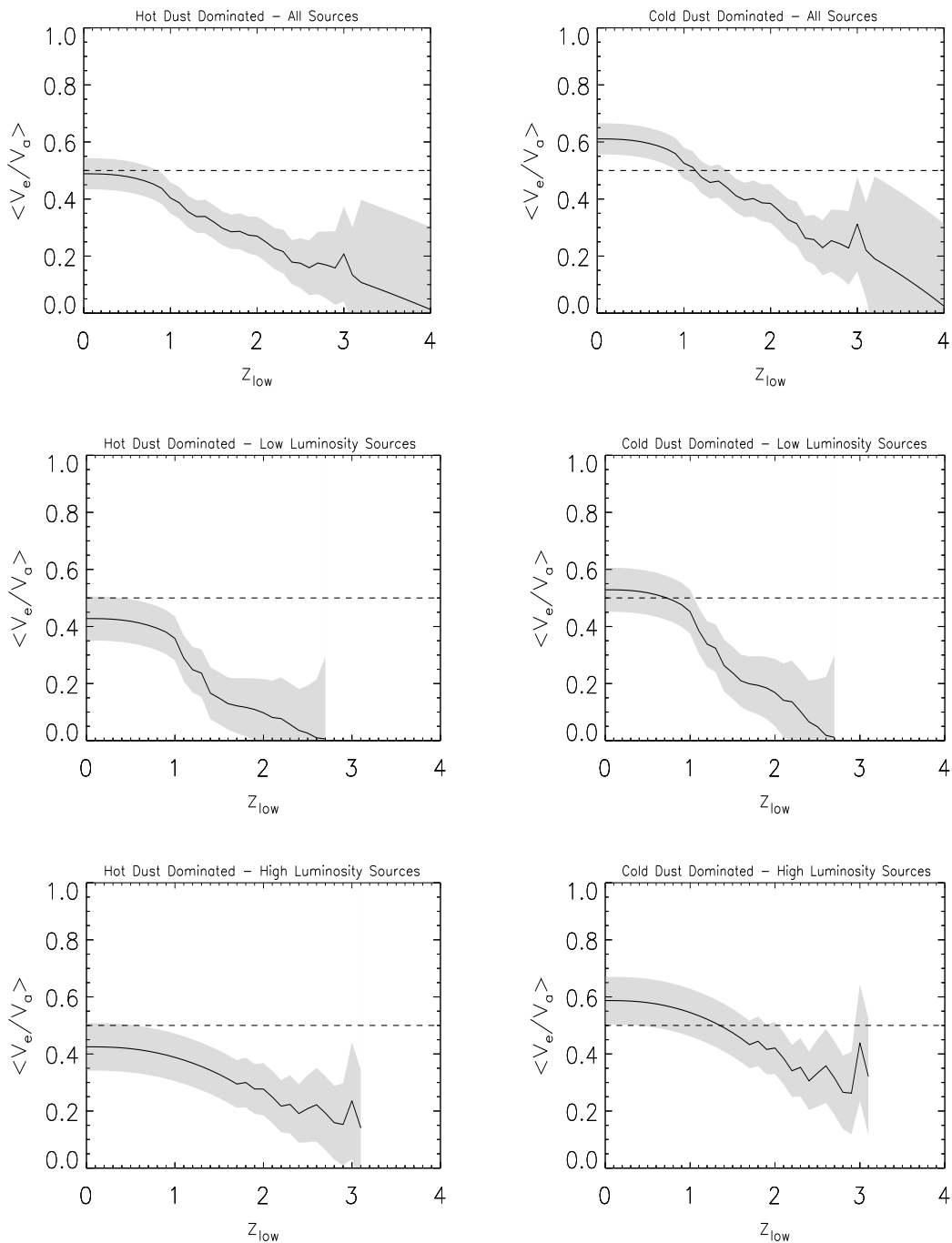}
\hfill}
  \caption{The distribution of the values of $\langle
  V_{e}/V_{a}\rangle$ with $z_{low}$ for the GOODS-N sample.  Figures
  in the left hand column are for hot SEDs and figures in the right
  hand column are for cold SEDs.  The sample is also separated into
  high and low luminosity sources.  The dashed line denotes the
  position of $\langle V_{e}/V_{a}\rangle=0.5$ on the graph, values
  above which imply a concentration of galaxies at higher redshifts
  and below which imply a concentration of galaxies at lower
  redshifts.  The grey shaded region shows the area enclosed by the
  1$\sigma$ error.}
  \label{fig:goodsveva}
\end{figure*}
 
Our results for the 38 SMGs of Wall, Pope \& Scott are shown in
figure~\ref{fig:goodsveva}.  We find good evidence for the existence
of a redshift cutoff at $z>1$ for the hot SED, and slightly weaker
evidence for a redshift cutoff at $z>2$ for the cold SED.  Dividing
the sample into separate populations of high and low luminosity
sources shows differences in the evolution of the two populations.
The low luminosity sources show much sharper redshift cutoffs whereas
the evidence for redshift cutoffs in the high luminosity sources is
far more marginal.  Thus we find evidence to support the conclusions
given in \cite{wall:2008}: there is a redshift cutoff for the sample
and that there is evidence for two separately evolving sub-populations

\begin{figure*}
\epsfxsize=150mm
{\hfill
\epsfbox{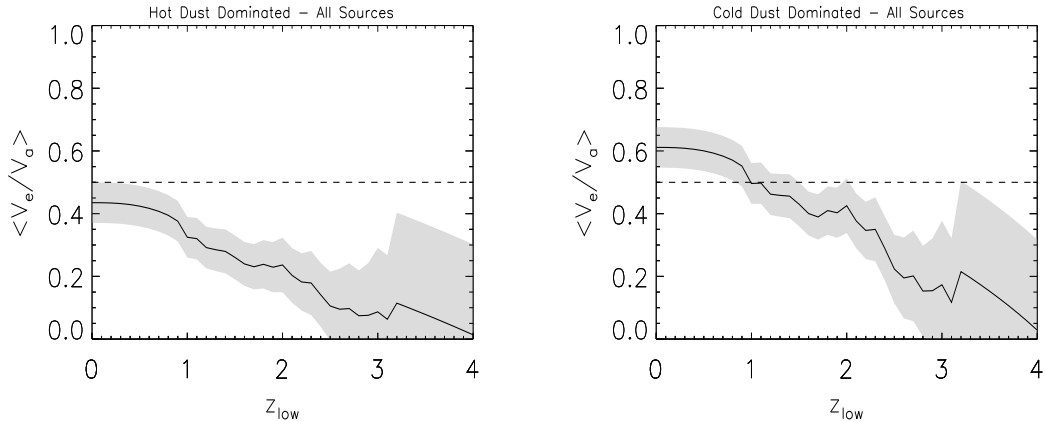}
\hfill}
  \caption{The distribution of the values of $\langle
  V_{e}/V_{a}\rangle$ with $z_{low}$ for the GOODS-N sample, using
  only the sources with reliable identifications.  Figures in the left
  hand column are for hot SEDs and figures in the right hand column
  are for cold SEDs.  The dashed line denotes the position of $\langle
  V_{e}/V_{a}\rangle=0.5$ on the graph, values above which imply a
  concentration of galaxies at higher redshifts and below which imply
  a concentration of galaxies at lower redshifts.  The grey shaded
  region shows the area enclosed by the 1$\sigma$ error.}
  \label{fig:goodsvevarobust}
\end{figure*}

An additional uncertainty about this results is that \citet{pope:2006}
claim that only 60\% of their identifications are reliable.  Therefore
we also performed the $\langle V_{e}/V_{a}\rangle$ analysis only on
sources with reliable identifications, the results of which are shown
in figure~\ref{fig:goodsvevarobust}.  Using these sources only, we
still find good evidence for a redshift cutoff at $z>1$ for the hot
SED, and some marginal evidence for a cutoff at $z>2$ for the cold
SED.  However, we are unable to find any clear evidence for two
separately evolving galaxy sub-populations, separated in luminosity,
as the sample size is too small.

\begin{figure*}
\epsfxsize=150mm
{\hfill
\epsfbox{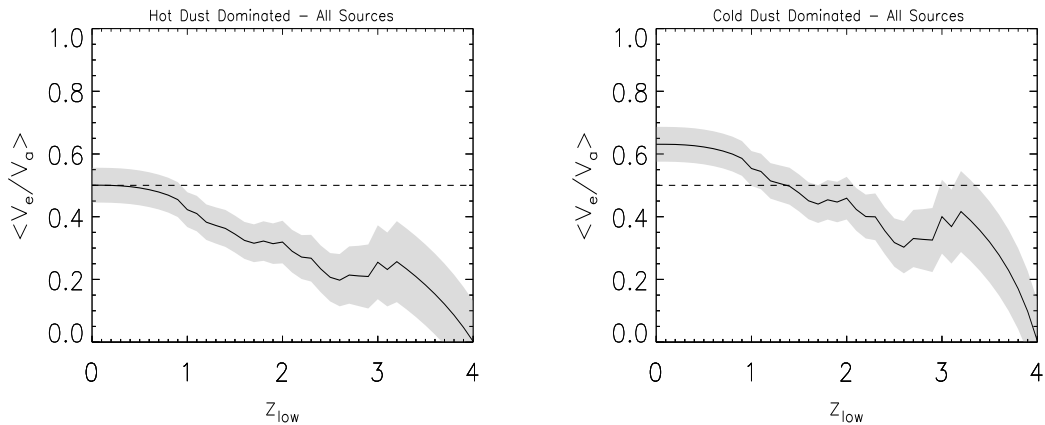}
\hfill}
  \caption{The distribution of the values of $\langle
  V_{e}/V_{a}\rangle$ with $z_{low}$ for the GOODS-N sample, where
  four (roughly half)of the unreliable identifications have been
  pushed to redshifts of $z=4$.  Figures in the left hand column are
  for hot SEDs and figures in the right hand column are
  for cold SEDs.  The dashed line denotes the position
  of $\langle V_{e}/V_{a}\rangle=0.5$ on the graph, values above which
  imply a concentration of galaxies at higher redshifts and below
  which imply a concentration of galaxies at lower redshifts.  The
  grey shaded region shows the area enclosed by the 1$\sigma$ error.}
  \label{fig:goodsveva_pushed}
\end{figure*}

However, by only taking into account the reliable identifications, we
are probably biased towards optically brighter galaxies and therefore
lower redshifts.  We further investigated the effect of the unreliable
identifications by putting four (roughly half) of the unreliable
identifications at $z = 4$ and repeating the analysis
(figure~\ref{fig:goodsveva_pushed}).  Doing this, we find that for hot
SEDs our results are largely unaffected, with a relatively clear
cutoff at redshifts higher than $z = 1$.  However for the cold SEDs we
find that our results are strongly affected, with no clear redshift
cutoff up to a redshift of z$\sim$3.

\begin{figure*}
\epsfxsize=150mm
{\hfill
\epsfbox{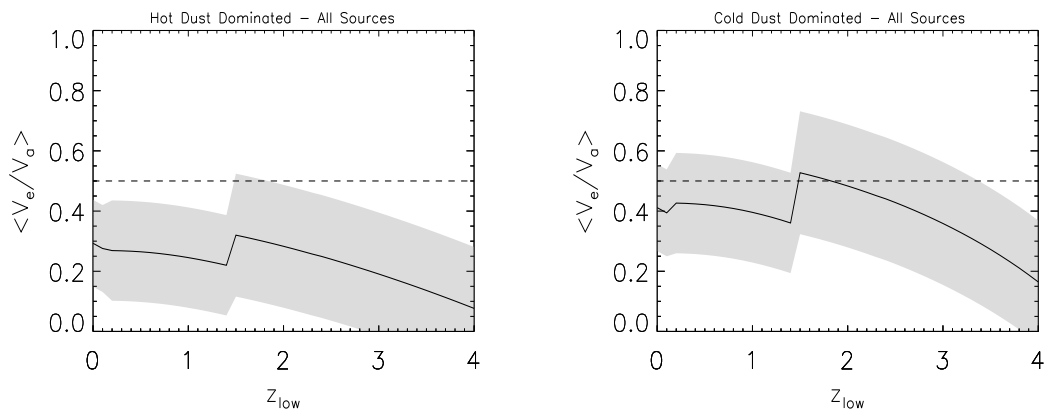}
\hfill}
\caption{The distribution of the values of $\langle
V_{e}/V_{a}\rangle$ with $z_{low}$ for the AzTEC sample.  The panels
on the left hand side uses the hot SED and the panels on the right
hand side uses the cold SED.  The dashed line denotes the position of
$\langle V_{e}/V_{a}\rangle=0.5$ on the graph, values above which
imply a concentration of galaxies at higher redshifts and below which
imply a concentration of galaxies at lower redshifts.  The grey shaded
region shows the area enclosed by the 1$\sigma$ error.}
\label{fig:aztecveva}
\end{figure*}

We also performed a banded $\langle V_{e}/V_{a}\rangle$ analysis on
our sample of AzTEC sources (excluding the AzTEC6 counterpart) the
results of which are shown in figure~\ref{fig:aztecveva}, but our
sample is too small to find any clear evidence of a redshift cutoff.

\section{CONCLUSIONS}

We give new Subaru, CFHT and IRAC photometry for a number of sources
in the AzTEC / COSMOS survey with accurate coordinates from SMA
imaging.  We have estimated photometric redshifts for four of the
seven galaxies in the sample.  We find a median redshift of
$z_{mean}\sim2.57$ and a maximum of $z_{max}=4.50$.  Of the sources in
the combined 850 $\mu$m surveys presented in \citet{chapman:2005},
\citet{pope:2006}, \citet{dye:2008} and \citet{clements:2008},
consisting of $\sim$200 sources, only one is at a redshift greater
than our two highest redshift sources.  This in addition to the fact
that we are unable to detect two of our sources in the optical bands
down to very faint magnitudes may indicate that 1.1~mm surveys are
more efficient at detecting very high-redshift sources than 850 $\mu$m
surveys.

Re-investigating the space density evolution of a sample of 38 GOODS-N
sources \citep{pope:2006,wall:2008} with more realistic SEDs we find a
redshift cutoff at $z\sim1$ if we assume a 'hot' SED and marginal
evidence for a cutoff at $z\sim2$ if we assume a 'cold' SED (in
reasonable agreement with Wall et al.).  Similar to \citet{wall:2008}
we also found evidence for two differently evolving sub-populations of
SMGs, separated in luminosity, with high luminosity sources showing a
less negative evolution.

We performed a similar test on the AzTEC sources but were unable to
draw any reliable conclusions as the sample is too small.  The GOODS-N
sample is also relatively small, and therefore any evidence for
redshift cutoffs and differently evolving sub-populations must be
treated with caution.  In order to harden our conclusions in general
we require larger surveys with accurate redshifts.  We would also need
surveys taken over larger areas of sky in order to take into account
the effects of cosmic variance.  Future, larger surveys (e.g. with
Herschel, SCUBA2) therefore will enable us to more robustly determine
the nature of the number density evolution of SMGs in the Universe.
 
\begin{flushleft}
{\bf Acknowledgements}
\end{flushleft}
G. Raymond, S. Eales and S. Dye acknowledge support from the Science
and Technologies Facilities Council.

Based on observations obtained with MegaPrime/MegaCam, a joint project
of CFHT and CEA/DAPNIA, at the Canada-France-Hawaii Telescope (CFHT)
which is operated by the National Research Council (NRC) of Canada,
the Institut National des Science de l'Univers of the Centre National
de la Recherche Scientifique (CNRS) of France, and the University of
Hawaii. This work is based in part on data products produced at
TERAPIX and the Canadian Astronomy Data Centre as part of the
Canada-France-Hawaii Telescope Legacy Survey, a collaborative project
of NRC and CNRS.

\label{lastpage}

\end{document}